# ELT HARMONI: Image Slicer Preliminary Design

Florence Laurent*[a], Didier Boudon[a], Johan Kosmalski[b], Magali Loupias[a], Guillaume Raffault[a], Alban Remillieux[a], Niranjan Thatte[c], Ian Bryson[d], Hermine Schnetler[d], Fraser Clarke[c], Matthias Tecza[c]

[a]Université de Lyon, Lyon, F-69003, France; Université Lyon 1, Observatoire de Lyon, 9 avenue Charles André, Saint-Genis Laval, F-69230, France; CNRS, UMR 5574, Centre de Recherche Astrophysique de Lyon; Ecole Normale Supérieure de Lyon, Lyon, F-69007, France ;
[b]European Southern Observatory, Karl-Schwarzschild-Str. 2, D-85748 Garching bei München, Germany
[c]University of Oxford, Department of Physics (Astrophysics), Denys Wilkinson Building, Keble Road, Oxford, Oxfordshire, OX1 3RH, UK
[d]Science and Technology Facilities Council, UK Astronomy Technology Centre, Royal Observatory, Edinburgh, Blackford Hill, Edinburgh, EH9 3HJ, UK

## ABSTRACT

Harmoni is the ELT's first light visible and near-infrared integral field spectrograph. It will provide four different spatial scales, ranging from coarse spaxels of $60 \times 30$ mas best suited for seeing limited observations, to 4 mas spaxels that Nyquist sample the diffraction limited point spread function of the ELT at near-infrared wavelengths. Each spaxel scale may be combined with eleven spectral settings, that provide a range of spectral resolving powers from R 3500 to R 20000 and instantaneous wavelength coverage spanning the 0.47 - 2.45 μm wavelength range of the instrument. The consortium consists of several institutes in Europe under leadership of Oxford University. Harmoni is starting its Final Design Phase after a Preliminary Design Phase in November, 2017.

The CRAL has the responsibility of the Integral Field Unit design linking the Preoptics to the 4 Spectrographs. It is composed of a field splitter associated with a relay system and an image slicer that create from a rectangular Field of View a very long (540mm) output slit for each spectrograph.

In this paper, the preliminary design and performances of Harmoni Image Slicer will be presented including image quality, pupil distortion and slit geometry. It has been designed by CRAL for Harmoni PDR in November, 2017. Special emphases will be put on straylight analysis and slice diffraction. The optimisation of the manufacturing and slit geometry will also be reported.

## 1. INTRODUCTION

HARMONI is a near-IR/optical integral field spectrograph. Four spaxel scales of 60x30, 20x20, 10x10 and 4x4mas give fields of view on the sky of 9.1x6.2", 4.1x3.0", 2.1x1.5" and 0.82x0.61" with increasing spatial resolution. Eleven spectral resolution settings of 3500, 7000 and 18000 are provided in all spatial scales, covering the wavelength range (not simultaneous) from 0.47 to 2.45 micrometres [1].

HARMONI will be located at the "side-looking" Nasmyth A3 port of the ELT. The instrument will work with seeing-limited (provided by the telescope), SCAO, and LTAO image qualities. SCAO uses a bright (R≤16) guide star to provide high Strehl image quality over a small fraction of the sky. LTAO combines laser guide stars with fainter natural guide stars to deliver diffraction-limited image quality in the JHK bands (~0.01") over a large fraction of the sky.

*florence.laurent@univ-lyon1.fr; phone +33 4 78 86 85 33; fax +33 4 78 86 83 86; http://cral.univ-lyon1.fr/

Light from the telescope is relayed to the instrument by a cooled relay system, to minimise additional thermal background. A user-selectable dichroic in front of the relay separates LGS light from NGS and science light to enable LTAO observations. The reflected laser light is redirected to a rotating LGSS system, which contains six laser guide star wavefront sensors. The dichroic is removed to allow visible light spectroscopy below 600nm. The 120-arcsecond NGS/science field transmitted by the dichroic is relayed to an 'up-looking' focus, which allows a gravity invariant rotating instrument configuration. Calibration light can be fed into the instrument via a deployable calibration system upstream of the relay.

The bulk of the instrument opto-mechanics, called Instrument Field Spectrograph (IFS), is housed in a single cryostat at ~140K (Figure 1). A pre-optics subsystem provides four selectable spatial pixel scales, in addition to other beam conditioning functions such as filters and pupil masks. The rectangular pre-optics output field is reformatted into the four linear slits by the IFU subsystem, each of which feeds one of four spectrograph subsystems. The spectrographs collimate, disperse, and focus the light onto detectors, providing a choice of spectral ranges and resolving powers. Each spectrograph has one camera for the NIR (0.8-2.5 micron) and up-to one for VIS (0.47-0.82 micron), each camera having a pair of 4k² detectors. The total number of VIS cameras will be determined by funding. The NIR detectors are further cooled down to ~40K.

Outside the cryostat, an SCAO wavefront sensor is fed by VIS light from a deployable dichroic covering the science field. A deployable module designed to "pre-process" the science light to optimise the high contrast performance of the instrument is mounted closely to the SCAO dichroic. Secondary Guiding and LTAO-NGS wavefront sensors (for seeing-limited and LTAO operation respectively) are mounted on an arm patrolling the entire 120-arcsecond technical field of view. All these wavefront sensing subsystems are mounted on top of (outside) the instrument cryostat, and so co-rotate with it on the instrument rotator. The instrument rotator also provides a services-wrap to carry services from the rotating section to the non-rotating facilities on the Nasmyth platform. The instrument and detector local control systems are mounted on/near the cryostat, to minimise cable lengths, while other parts of the instrument local control system is mounted on the Nasmyth platform to minimise moving mass/volume. High level control and data handling software are located remotely in a control room (Figure 1).

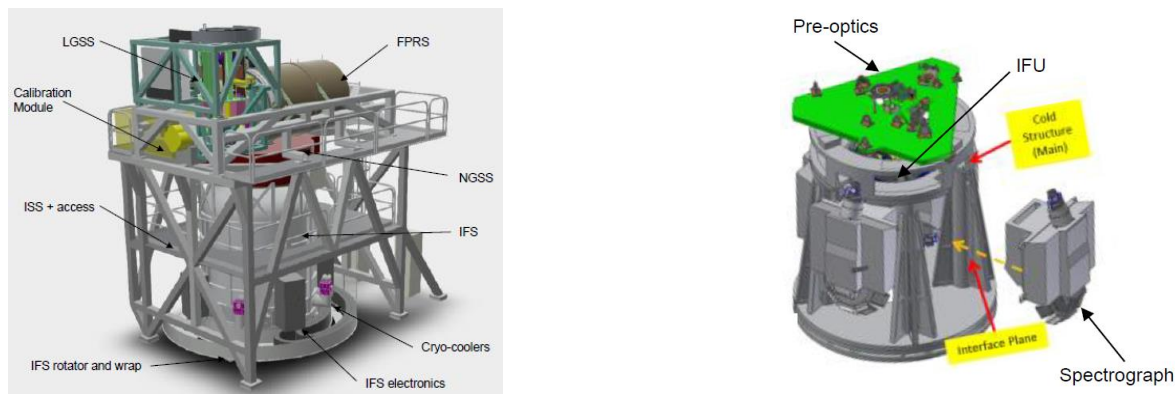

Figure 1: Left: CAD layout of HARMONI on the Nasmyth platform, showing all the major sub-systems, the support structure, rotator and cable wrap. Right: CAD layout of HARMONI Integral Field Spectrograph where the IFU is located between the Pre-optics and the Spectrograph.

The Integral Field Unit (IFU) Sub-System which is under CRAL responsibility is part of the HARMONI Integral Field Spectrograph inside the rotating cryostat [2]. It is in charge of dividing the Field of View, reshaping it to provide 4 output slits to the 4 spectrographs. Its input is the rectangular field of view from the Pre-Optics Sub-System, it cuts it into 8 sub-field of views through the Splitting and Relay Module (SRM). After magnification, pairs of sub-fields are entering into two Image Slicer Units (ISU). In each of the ISU pair, the field is sliced in 2x38 mini slits, and rearranged along an output slit of 76 mini-slits, to feed the spectrographs.

This paper focuses on the Image Slicer preliminary design proposed for Harmoni Preliminary Design Review in November, 2017. Section 2 and 3 of this paper introduce the ISM optical and mechanical designs. The optimisation of the manufacturing and slit geometry is also reported. The key performances coming from the optical design and associated analysis are described in section 4. Special emphases are put on straylight analysis taking into account mirror roughness and slice diffraction. Finally, section 5, compiles the conclusions and future developments.

## 2. ISM OPTICAL DESCRIPTION

### 2.1 Key Requirements

The Image Slicer Module is made up of 8 Image Slicer Unit (ISU) the function of which, are:
- Slice the sub-FoV to reach the spatial resolution required.
- Rearrange the FoV to produce an output slit to feed the spectrograph. That means that the IFU shall produce four output slits.
- From a telecentric input, produce a telecentric output with the superimposition of all the pupils (ie 76 sub-pupils).

The key requirements are the following:
- The input fastest f/ratio is 93 with an input field size of 51x38mm for each ISU.
- Image quality: The IFU wavefront error shall be lower than 300nm RMS for 60x30 spatial scale. So, the ISM wavefront error shall be lower than 200nm RMS for 60x30 spatial scale.
- Output Slit geometry. The IFU shall produce four output slits. Each slit will lower than 541.8mm long. Within each slit, all mini-slit shall be within ±4.5mm (in the spectral dispersion direction) of a nominal line. Each slice of the IFU shall be re-imaged to 130±3µm along spectral direction and with a length of ≥6.63mm in the exit focal plane of the IFU.
- Output Pupil positioning. The IFU shall generate an exit pupil with a distance >±20m and the exit pupil centres of all field points shall be concentric with the nominal exit pupil, to within 2.5% of the pupil diameter of the fastest f/ratio.

These main requirements will be analysed in section 4.

### 2.2 Optical Description

The ISM is composed of 8 Image Slicer Units (ISU). The ISU is either an ISU_TypeA or an ISU_TypeB which are slightly different. A general overview is given in Figure 2 and Figure 3.

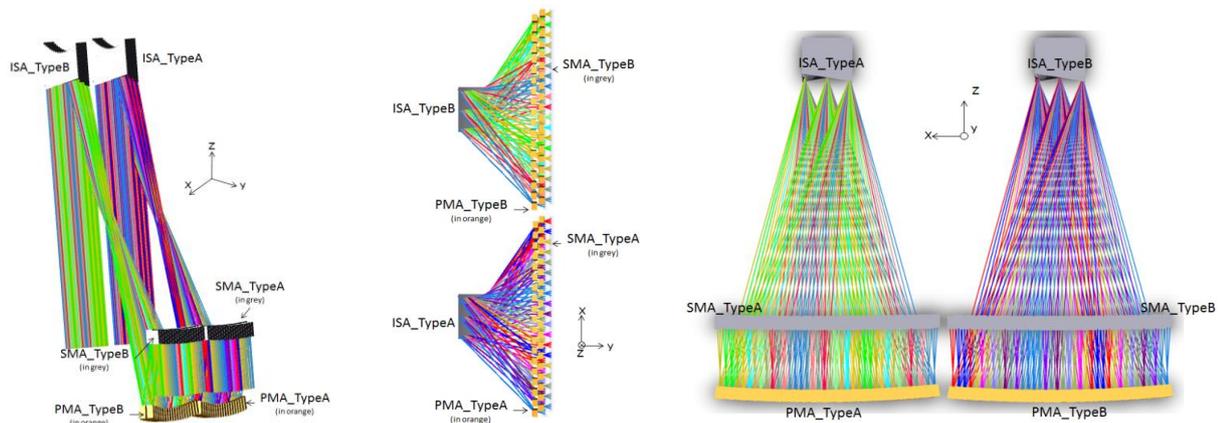

Figure 2: ISU_TypeA & ISU_TypeB XY/XZ/YZ views which provides one output slit of 76 mini-slits for one spectrograph

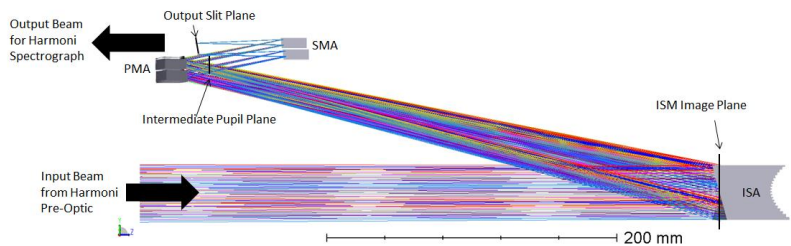

Figure 3: ISU pupil and image planes, input and output beams

Each ISU is composed of 3 elements (ISA, PMA and SMA).
- Image Slicer Assembly (ISA): This assembly is composed of 1 stack of 38 thin Mirrors (commonly called slices) named ISA_TypeA or ISA_TypeB. The design of ISA is optimised to fit the process described in [3], to be able to manufacture the 8 ISA blocks in only 4 parent polishing parabolas of diameter 330mm (§ 2.3). That means that 2 stacks of 38 thin Mirrors come from the same Parabolic Mirror with R=800mm. Each slice (51x1mm) has a different off-axis angle and cuts the entrance Field of View into thin, narrow stripes, redirecting the beams towards the Pupil Mirror Assembly (PMA) and imaging the Telescope Pupil in an intermediate pupil plane located between PMA and Slit Mirror Assembly (SMA). Each Slice is associated with one mirror of the PMA and one of the SMA. The position of the slice to ±50µm in the polishing parabola drives the tolerance of manufacturing of the slices angle to ±13 arcsec. After polishing, the slices optically contacted inside the polishing parabola are disassembled, ordered on each stack, and optically contacted again, through interferometric control on a reference surface. This process allows to reach the high performance requested in terms of angular accuracy, as demonstrated in [4] and [5].
- A Pupil Mirror Assembly (PMA): This assembly is composed of 38 tilted flat mirrors (9x7mm) called PMA_TypeA or PMA_TypeB. They are located ~20mm before the intermediate pupil plane created by the ISA and they are used to compensate the quite high angle of the beam coming from the ISA towards the SMA. Thus, the SMA can work with acceptable angle of incidence needed to meet the image and pupil quality of the whole system. The size of the intermediate pupil is 7.2mm and the footprints on the PMA arrangement of mirrors are in the Figure 4.
- A Slit Mirror Assembly (SMA): This assembly is composed of 38 off-axis spherical mirrors (13x7mm) called SMA_TypeA or SMA_TypeB. The SMA implements three functions namely: (1) deflecting the optical beams from the PMA so that they are parallel one to each other, (2) creating a demagnified image of its associated slice at the entrance of the Spectrograph, and (3) re-imaging all the images of the telescope pupil at a common location that will be the entrance pupil of the Harmoni spectrograph. Each mirror has a different radius of curvature (from 104.6 to 108.1mm) to compensate for the optical path difference between the different paths. The tolerance of the curvature center position is also of ±50µm, but the main constraint is the arrangement of the array. The 1 mm space between the mirrors is the constraint. The footprints on the SMA arrangement of mirrors are in the Figure 4.

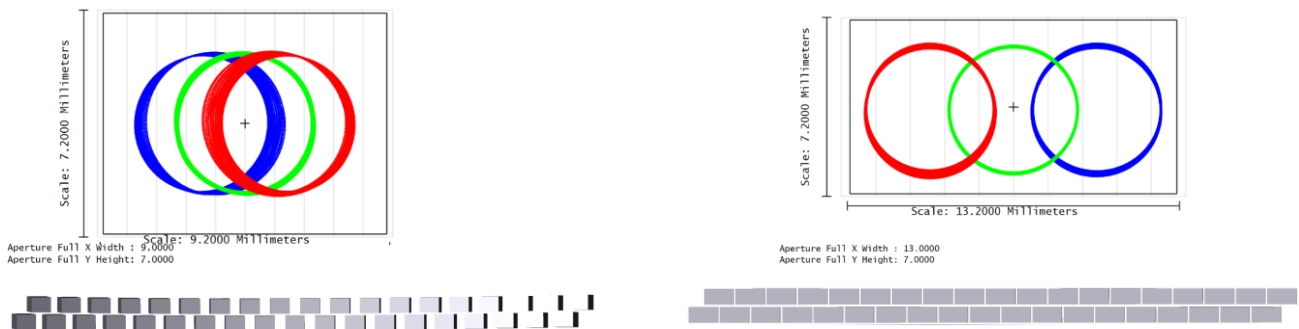

Figure 4: Left: Pupil mirror array footprint and arrangement. Right: Slit mirror array footprint and arrangement

At the exit of 2 Image Slicer Units (ISU_TypeA & ISU_TypeB), two rectangular FoVs have been transformed into only one long output slit made of 76 mini-slits. Note the need to have the following combination ISA_TypeA&ISA_TypeB / PMA_TypeA&PMA_TypeB / SMA_TypeA&SMA_TypeB to create one output slit for one spectrograph.

### 2.3 Manufacturing Optimisation

As described in section 2.2, there are 8 ISA blocks of 38 slices each. For the Harmoni PDR, the goal was to optimize the slice arrangement onto the parent polishing parabola in order to decrease the manufacturing cost.

The image slicer polishing parabola arrangement was investigated for few cases:
- Case_1: Nominal - ISA with Phase A design
- Case_2: ISA with S-Shape
- Case_3: ISA with 26.5mm x-decentring
- Case_4: ISA with different y-decentring

In all cases, ISM performances and polishing parabola optimization are compared.

- Case_1: Nominal - ISA with Phase A design

For the Phase A design, the 8 ISA blocks are identical that involves 8 parent polishing parabolas of diameter 320mm.

- Case_2: ISA with S-Shape

The design is the same as Phase A, expect that the two ISA are not identical for the PMA arrangement. First ISA distributes light from the right up to left, second from the left up to right. There are 4 identical ISU. The positioning of each slice onto the same optical parabola is defined in Figure 5. That involves 4 parent polishing parabolas of diameter 320mm and a small parent parabola should be manufactured in order to manufacture the central slices.

- Case_3: ISA with 26.5mm x-decentring

The design is the same as Phase A, expect that the two ISA are moved from 26.5mm along x-axis from each other. There are 4 identical ISU. The positioning of each slice onto the same optical parabola is defined in Figure 5. That involves 4 parent polishing parabolas of diameter 400mm.

- Case_4: ISA with different y-decentring

The design is the same as Phase A, expect that the two ISA are different in term of slice y-decentring. The consequence is that the two PMA rows are shifted along y-axis of a different value. There are still 4 identical ISU. The positioning of each slice onto the same optical parabola is defined in Figure 5. That involves 4 parent polishing parabolas of diameter 330mm.

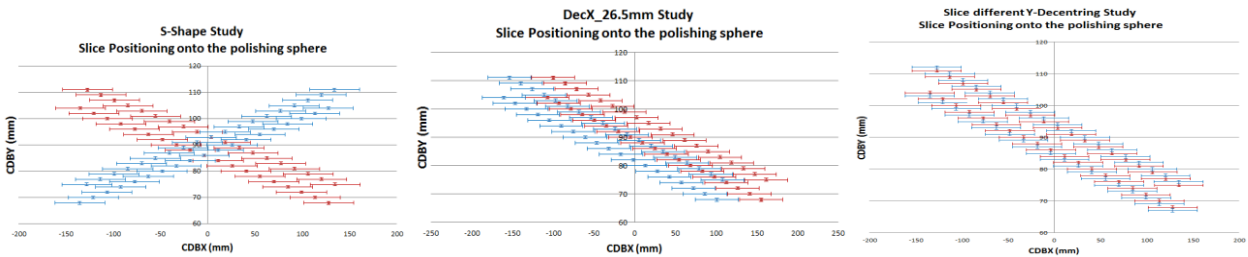

Figure 5: Slice positioning onto one polishing parabola for two ISA. Left: Case_2. Middle: Case_3. Right: Case_4

In all cases, ISM performances and polishing parabola optimisation are compared. A trade-off was made in summarizing the pros and cons of each case (Table 1). For PDR, the design of ISA where the y-axis is decentered has been chosen because it presents the best arrangement onto the polishing parabola with compliant ISM performances.

Table 1: Trade-off to optimize slicer manufacturing

| Item | ISM Performances | Number of Polishing Parabola | ISA/PMA/SMA identical |
|---|---|---|---|
| Case_1: Nominal - ISA with Phase A design | Compliant | 8 | All identical |
| Case_2: ISA with S-Shape | Compliant | 4 and 4 small ones | A/B different |
| Case_3: ISA with 26.5mm x-decentring | Compliant | 4 but with larger polishing parabola | A/B different |
| Case_4: ISA with different y-decentring | Compliant | 4 | A/B different |

## 2.4 Output Slit Geometry Optimisation

The output slit geometry was investigated for two cases:
- Case_1: Staggered slit with tilted focal plane
- Case_2: Staggered and curved slit with curved focal plane

In all cases, ISM performances and manufacturing constraints are compared.

- Case_1: Staggered slit with tilted focal plane

The 8mm slit stagger ie ±4mm Y field position with tilted focal plane around the spatial direction of 10.1° was designed. The performances are compliant with requirements and manufacturing constraints are acceptable (§4).

- Case_2: Staggered and curved slit with curved focal plane

The staggered and curved slit with a curved focal plane was designed. That means the focal plane is curved in xy and xz directions (Figure 6). This Case_2 presents degraded performances. The image quality and pupil oversizing are twice worse than Case_1. Moreover, this Case_2 has more difficult manufacturing constraints because all optomechanical elements are different. To improve performances, the SMA could be 38 off-axis toroidal mirrors. In this case, the performances are roughly similar to Case_1 but with toroidal mirrors, the manufacturing is not feasible with classical polishing.

By design, the output slit geometry could be whatever you dream. But, the manufacturing becomes unfeasible at an affordable cost. For PDR, the output slit geometry of the Case_1 has been chosen for cost reason (§4.2).

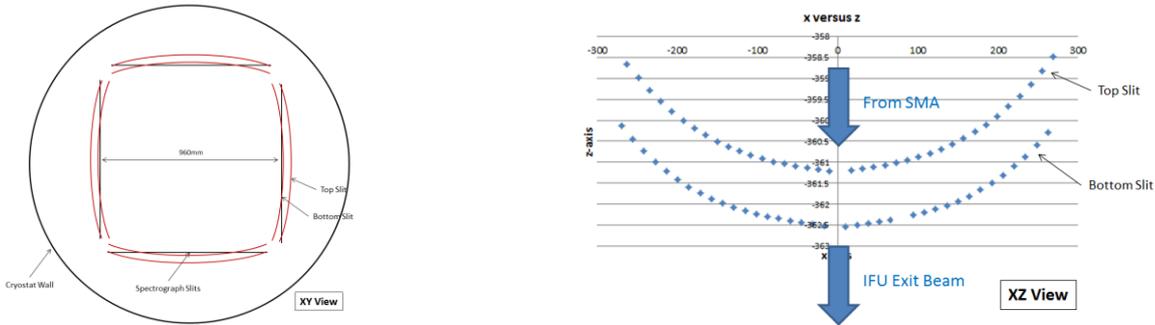

Figure 6: Illustration of Case_2 - Staggered and curved slit with curved focal plane

## 3. ISM MECHANICAL DESCRIPTION

The Slicer Mechanical Support (SMS) aims at positioning accurately the pair of Image Slicer Assembly (ISA), Pupil Mirror Assembly (PMA), Slit Mirror Assembly (SMA) and output Slit Baffle Assembly (SBA). There are 4 identical supports (Figure 7). The whole Slicer Mechanical Support is made of 7075 aluminum alloy, which ensures a good rigidity, together with a reduced mass. Some support-legs link the horizontal and vertical plates to strengthen the structure. The center of gravity is exactly located at the center of each module, at the interface level, which provides a good stability during assembly. The mass of each SMS is 35 Kg including optics. For FDR, the increase of the clearance of the output beam and the adaptation of the SMA position in the optical design giving more margins for the mechanical encumbrance are foreseen.

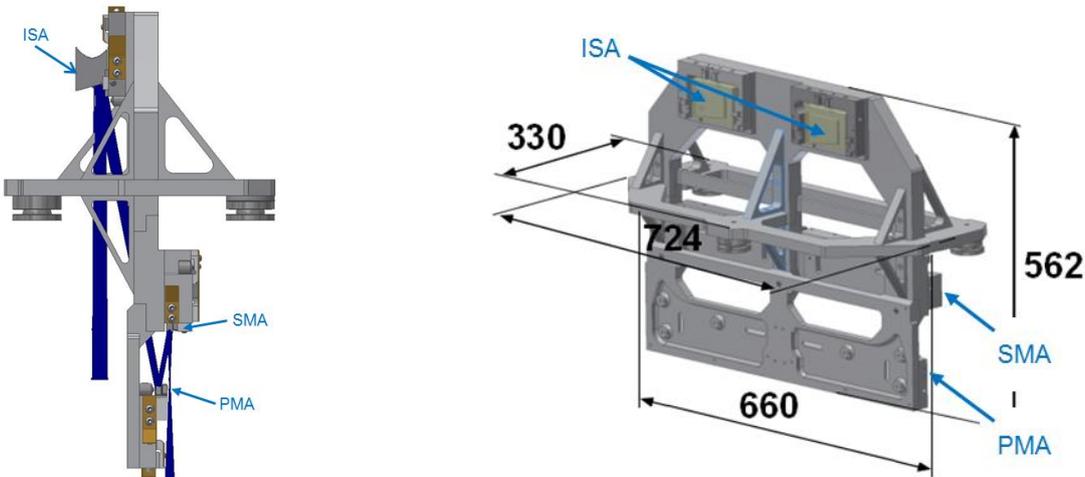

Figure 7: SMS mechanical design. Numbers are in mm

The SMS is composed of 3 pieces (for manufacturing reasons), which are pinned and screwed together as illustrated onto the Figure 8. The SMS could be realized in one block but the integration would be less comfortable. Both cases will be analyzed. The interface between the SMS and the IFU uses a kinematic system. The point of the kinematic system, which is the reference, is located between the 2 Image Slicer Assemblies. Before integration, each SMS will be pre-adjusted on a bench using the same interface.

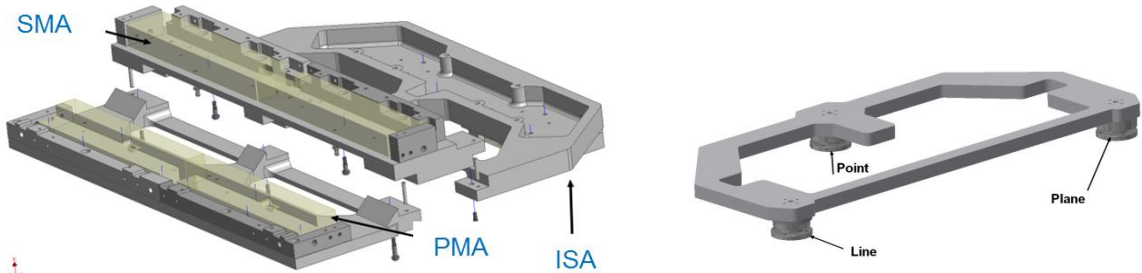

Figure 8: Left: Slicer Mechanical Support. Right: Interface to IFU flange

All the optical components of ISA, PMA and SMA are positioned in an isostatic way in their mount. Along X and Z axis, 3 shafts allow an accurate positioning of the optics. The shafts are machined at the right dimension and they take into account the differential expansion of the different materials (optics in Zerodur / mechanical mounts in Aluminum alloy). Along Y axis, the positioning is ensured by 3 pins, which can also be re-machined if necessary (Figure 9).

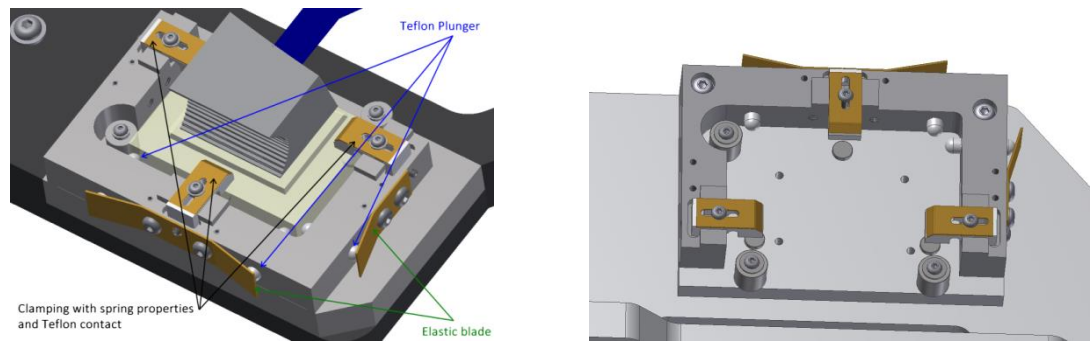

Figure 9: ISA positioning mechanical design

The clamping is made by chrysocale clamps, protected with polymer (PCTFE). This allows a safe contact with the optical components and a good sliding, necessary for the thermal expansion. An elastic system, using elastic blades and teflon plungers, ensures a constant pressure of the optical component against the isostatic points, whatever the temperature, between room temperature and 140K.

## 4. ISM PERFORMANCES

These analyses onto ISM are performed on the following assumptions:
- In the largest spatial scale ie. 60x30mas which represents the worst case. Only the diffraction analyses takes into account smallest scales.
- In order to show the performances along the output slit, 3 representative fields along x-axis are selected per configuration ie per mini-slit. That means 114 fields. To simplify some graphs, where there is no difference between central and extremity fields, only the central field is shown.
- As the ISM is an all reflective design, all requirements are wavelength independent Only the diffraction analyses takes into account the wavelength variation (0.47, 0.8, 2.45 microns).

## 4.1 Image Quality

The image quality analysis is performed by Zemax and does not include the diffraction. The Figure 10 shows the wavefront map at the output slit plane. The optical performance is compliant with requirements with an average RMS wavefront error of 140nm. That gives margins for MAIT phase.

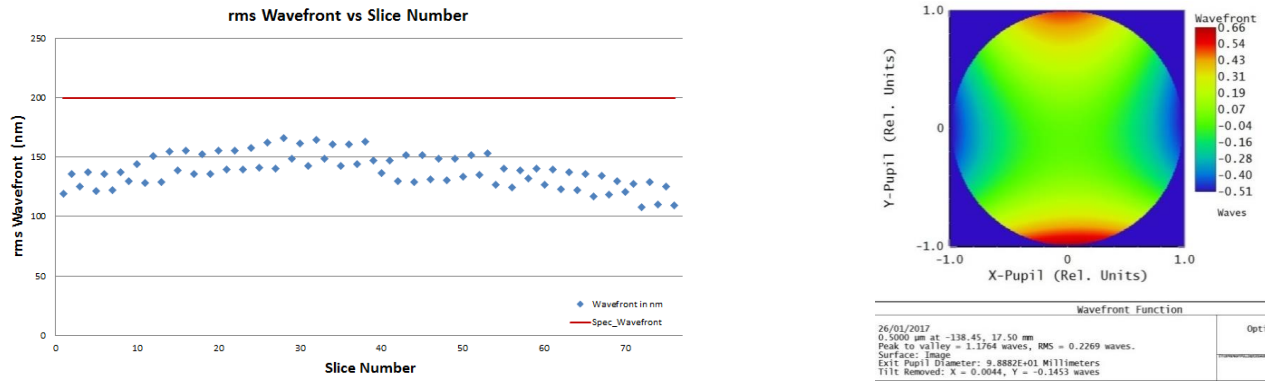

Figure 10: Left: Wavefront value for all slices for central field. Right: Mean Wavefront Map in output slit plane

## 4.2 Ouput Slit Geometry

The output slit presents 2 straight lines of 38 mini-slits staggered with ±4.0mm in the y-axis spectral direction without any bending along the spectral direction at the spectrograph object plane (Figure 11). The ouput slit is 538.6mm long, including a 15.3mm gap for detector spacing and some extra spacing for alignment tolerances. The gaps in X (on CCD or inter mini slit gaps or margins on the slit sides) can be easily implemented with other values in the next phase if necessary. Each line has different focus position: The bottom and the top one are focused respectively at -0.747mm and +0.683mm from the output slit nominal position (Figure 11). The exit flare angle is at 0° around x&y axes.

As detailed in section 2.4, each line could be slightly pre-shaped to reduce the effect of spectra bending by the grating. Nevertheless, as the bending depends on which grating is being used, the shape can only be a compromise between all grating configurations. Note that, the more the pre-shaped will be "exotic", the more the SMA manufacturing will be difficult or impossible. For Harmoni, a staggered slit with tilted focal plane has been chosen. A zoom of the output slit shape is presented onto Figure 11.

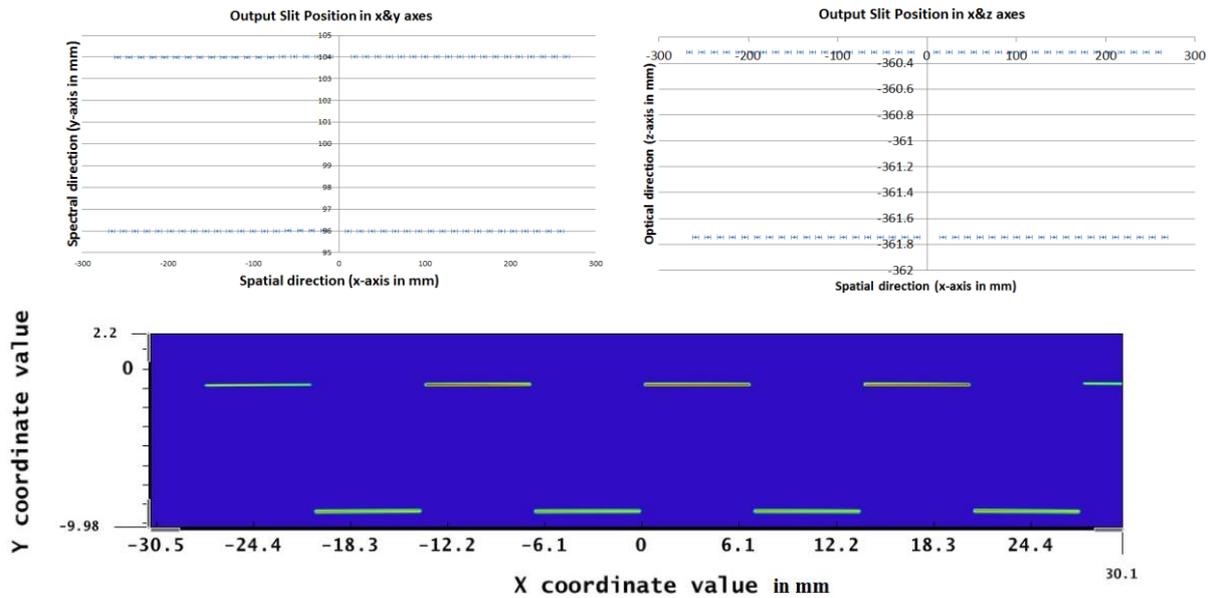

Figure 11: Output slit at the level of the entrance of the Spectrograph. Upper Left: x&y axes, that illustrates the y-positioning of the slit. Upper Right: x&z axes. Bottom: Zoom of a few slices

The mini-slit height and length are computed with the geometrical magnification computed by Zemax without diffraction along x&y axes. The maximum magnification variation is ± 0.2% along the slice between the 76 mini-slits due to the SMA and PMA curvature and alignment variation. The mini-slit length is compliant with the requirement of < 6.63mm (Figure 12). The maximum magnification variation is ± 1.2% in the height of slice. The mini-slit height is partially compliant with requirement but it can be easily compliant with requirements in the next phase without performance degradations (Figure 12).

Each mini-slit in the output slit is separated in the spatial direction by 193μm in average from its neighbors.

Note that each mini-slit is tilted around the optical axis (from -2° up to +2.3°) - Figure 13. This tilt is higher at the extremities and at the centre of the output slit.

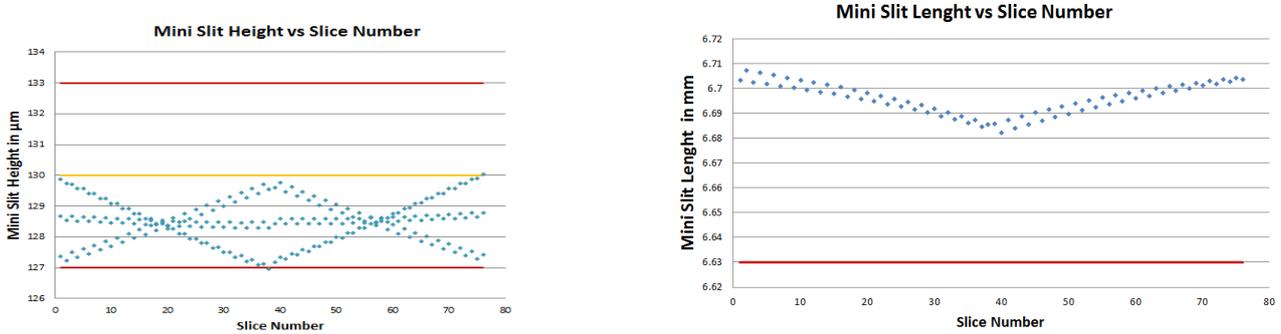

Figure 12: Left: Mini-slit height for all slices and 3 points along the slice. Right: Mini-slit length for all slices

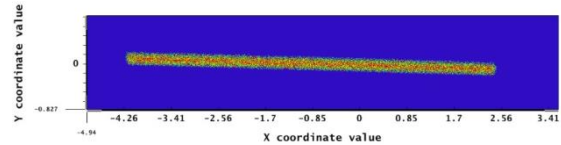

Figure 13: Mini-slit tilt around optical

## 4.3 Output Pupil Positioning

The exit pupil corresponds to the overlapping of the 76 sub-pupils coming from the 76 slices. This overlapping might be reduced and as the pupils get smaller, lines density and efficiency in-homogeneities of the grating as well as mirror coatings inhomogeneity might be an issue.

Onto the 76 mini-slit, the minimum pupil position along z-axis is higher than 550m which is compliant with requirement being superior to 20m (Figure 14).

The centre pupil analysis is performed in Zemax using a paraxial lens of focal length 1500mm and computing expected pupil sizes and centres at its focus. The pupil degradation for the other spatial scales can not introduce any vignetting in the spectrograph due to a smaller pupil.

The exit pupil centres of all field points are not concentric with the nominal exit pupil along y-axis within ±0.1% of the pupil diameter. The total exit pupil diameter has to take into account both the exit pupil position variation along x&y axes and the error of 2% on the exit f-number. As a consequence the total exit pupil diameter is 128.2mm +0.1% in X and +1.2% in Y which is largely smaller than the pupil wander (Figure 15).

In conclusion, the exit pupil is compliant with requirements. Nevertheless, when the Image Slicer Module is combined with the Relay System Module, the requirement becomes very stringent. There is no margin for MAIT. These results are detailed in the paper [2].

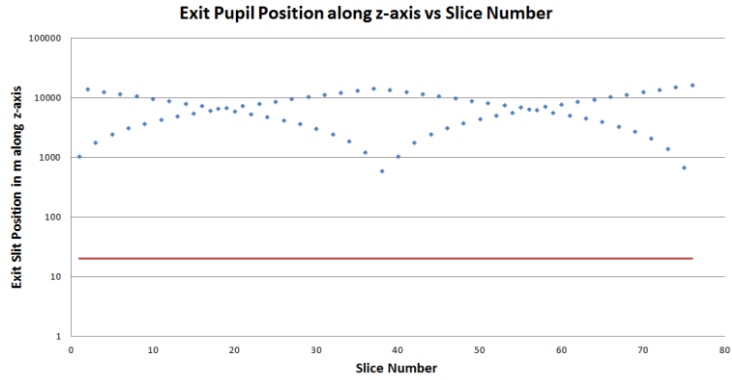

Figure 14: Exit pupil position in m along z-axis.

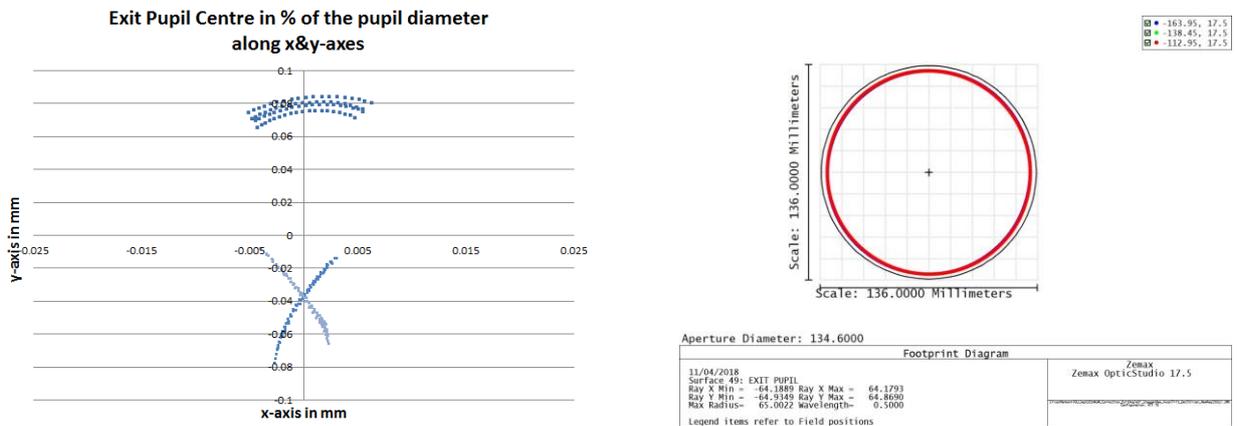

Figure 15: Exit pupil position. Left: Exit pupil centre for 3 points on the slice along x&y axes. Right: Exit pupil corresponding to the overlapping of the 76 sub-pupils

## 4.4 Specular Reflection

Because of the optical design of the slicer, the underside of a slice can receive light from the active face of the bottom slice and create ghost images. It represents a light loss of 2.83% (blue rays in Figure 16). This light is not directed onto the output slit plane. To reduce that, the optical design of the slicer could be easily changed. This will be done during next phase if necessary.

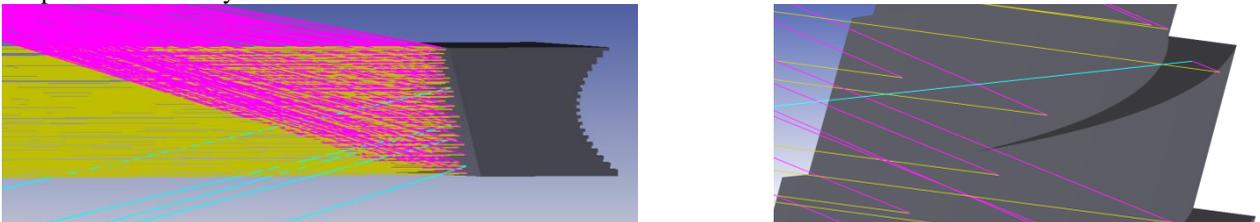

Figure 16: Specular reflection on ISA

## 4.5 Straylight

In order to perform realistic analysis considering the roughness of the active surfaces, an ABg scattering model in zemax is used with the following parameters: A=6.7E-4 | B=0.001 | g=1.5 based on a paper [6]. This corresponds to a roughness of 30Å which is pessimistic with classical polishing.

Only one slice is illuminated and ABg model is applied separately to ISA, PMA, SMA and finally all of them. 5000000 rays are launched and recorded onto a detector in the output slit plane to evaluate light contamination. To simulate the mechanics obscuration, a rectangular baffle is put behind the PMA. The Figure 17 presents the output slit plane in log-15 scale when only the slicer is scattering. The red mini slit is the specular reflection of the illuminated slice. The other mini slits are ghosts images. When an ABg model is applied onto the 3 components, the straylight is around 0.3%. The maximal value on a slit pixel of 65μm side can lose up to 0.1% (Figure 17).

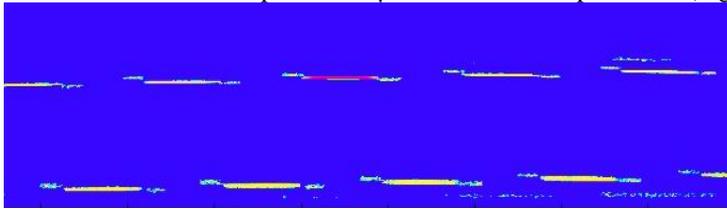

| Scattering Component | % of Straylight | Pixel Max Loss (%) |
|---|---|---|
| ISA | 0.02 | 0.03 |
| PMA | 0.09 | 0.03 |
| SMA | 0.20 | 0.09 |
| ALL | 0.30 | 0.08 |

Figure 17: Left: View of the output slit plane (log-15 scale), only the slicer is scattering, the red mini slit is the specular reflection of the illuminated slice. The other mini slits are ghosts images. Right: Table with percentage of straylight when components are scattering

To reduce the amount of straylight at the output slit plane, a mask with mini-slit holes 300% oversized (~400microns height) is used as a baffle. With this mask, we can reduce by 100 the amount of straylight at the output slit plane (Figure 18).

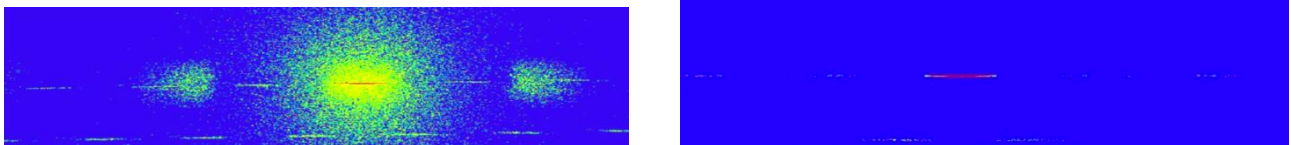

Figure 18: View of the output slit plane in log-15 scale when all components are scattering. Left: Without mask. Right: With mask

At PDR level, the straylight seems acceptable without mask with pessimistic assumptions. The mask is not yet implemented as it requires to modify slightly the optical design of the Image Slicer and because the analysis shows acceptable results. This will be done during next phase if necessary.

### 4.6 Diffraction Analysis

For the different spatial scales of Harmoni, the PSF imaged on the ISA can be significantly larger than the slice height. This has 2 effects:
- The PSF is sliced into several parts that fall at different location onto the Detector. This effect is similar to the case of a PSF smaller than the slice aperture but decentered with respect to the slice aperture. It is not an issue as an appropriate reconstruction of your image can give access to your real PSF.
- The pupil is diffracted by the slice aperture and elongated in the direction perpendicular to the Slice width according to the diffraction theory. In this case, some light can contaminate the neighbor PMA and create an unwanted image of the PSF. That is this effect which is analysed here.

The calculation of the Airy spot size can be found on the Table 2. For the Spatial Scale 60x30mas, the Airy central peak is smaller than the slice height (ie 1mm), so slice diffraction is negligible. The diffraction study has been carried out for the 3 smallest cases where the slice diffraction becomes significant. As PSF size increases with wavelength, the worst case at 2.45μm is used for this analysis. Moreover, the PSF is positioned on the extremity of the slice to evaluate the worst impact on the PMA.

Table 2: Airy disk diameter for the 4 Harmoni spatial scales

| Item | Scale: 60x30 | Scale: 20x20 | Scale: 10x10 | Scale: 4x4 |
|---|---|---|---|---|
| f/D along spatial direction (x-axis) | 90 | 135 | 270 | 674 |
| f/D along spatial direction (y-axis) | 90 | 269 | 540 | 1348 |
| Airy Ellipse diameter_x (mm) | 0.537 | 0.806 | 1.611 | 4.028 |
| Airy Ellipse diameter_y (mm) | 0.537 | 1.611 | 3.223 | 8.057 |

The Figure 19 presents one PSF on the slice for the 4x4 scale (left) and one diffracted pupil image at the level of PMA for the 20x20 scale (black rectangles are the PMA mirrors position). This study computes the contamination due to diffraction for the 3 smallest scales onto the neighbor PMA (Table 3).

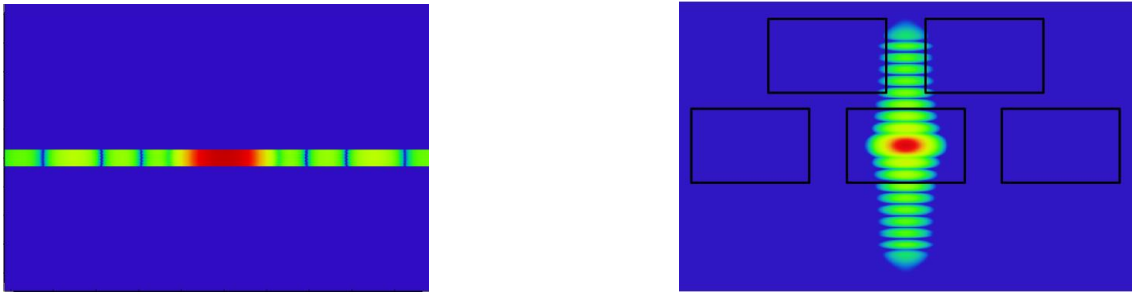

Figure 19: Left: PSF on the slice for the 4x4scale. Right: Diffracted pupil image at the level of PMA for the 20x20 scale (black rectangles are the PMA mirrors position)

The amount of light on the upper PMA is negligible compared to the straylight analysis results. The maximal value of 0.05% is reached for the spatial scale 20x20 (Table 3).

Table 3: Contamination due to diffraction for the 3 smallest scales

| **Item** | **Scale: 20x20** | **Scale: 10x10** | **Scale: 4x4** |
|---|---|---|---|
| Power on the central slice | 61% (PSF cut into 3 parts) | 55% (PSF cut into 5 parts) | 21% (PSF cut into 8 parts) |
| Light on the upper PMA (log-10 scale) | 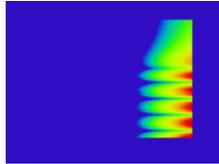 | 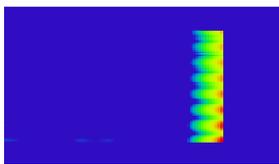 | 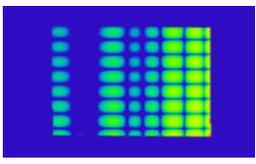 |
| Power on the upper PMA | 0.05% | 0.004% | 0.003% |

At PDR level, the diffraction analysis seems acceptable.

### 4.7 Sensitivity Analysis

The alignment sensitivity analysis is done using Zemax. The approach to the analysis is as follows:
- Beginning with the nominal optical design, each optical group is individually perturbed by a prescribed amount in each degree of freedom (DoF). For most groups, the perturbations that are modelled are lateral displacement (x, y), angular rotation (θx, θy, θz), and longitudinal displacement (z).
- The sensitivity analysis is performed on 4 optical groups (ISU, ISA, PMA and SMA).
- Three main criteria in the system are evaluated on the output slit plane:
    - The output slit position
    - The exit pupil position and size
    - The RMS spot radius (RSCE) degradation (faster in zemax to compute RSCE rather than wavefront as specified).

Dividing the change by the magnitude of the perturbation yields the Alignment Sensitivity for optics in that DoF. The coefficients are given in mm at the output slit per mm of the mirror or mm per arcmin. For example, decentering the ISU of 1 mm resulted in an average increase in the system slit positioning of 1.13mm, and the resultant Alignment Sensitivity is therefore 1.13mm/1mm. No compensator is used to compute the sensitivity coefficients.

The main contributors for:
- The image shift are the ISU (x, y, θx) and SMA (x, y, θx) - Table 4.
- The exit pupil degradation are ISU (θx, θy), ISA (x, y, θx, θy, R), and SMA (x, y) - Table 4

- The image quality degradation are ISU (z, θx), ISA (y, θy), and SMA (y, z, R)
- The sensitivity in Image quality degradation is much more tolerant than the pupil and slit position variation.

Table 4: ISU Sensitivity coefficients. Left: For Mini slit positioning. Right: For exit pupil positioning

| | | ISU | | | | | |
|---|---|---|---|---|---|---|---|
| | | x | y | z | θx | θy | θz |
| Mini-Slit position (mm) | x | 1.13 | 0.00 | 0.00 | 0.00 | 0.00 | 0.00 |
| | y | 0.00 | 1.13 | 0.00 | 0.00 | 0.00 | 0.00 |
| | z | 0.00 | 0.17 | 0.02 | 0.03 | 0.00 | 0.00 |
| | | ISA | | | | | |
| | x | 0.01 | 0.00 | 0.00 | 0.00 | 0.00 | 0.00 |
| | y | 0.00 | 0.01 | 0.03 | 0.00 | 0.00 | 0.00 |
| | z | 0.00 | 0.00 | 0.00 | 0.00 | 0.00 | 0.00 |
| | | PMA | | | | | |
| | x | 0.00 | 0.00 | 0.00 | 0.00 | 0.03 | 0.00 |
| | y | 0.00 | 0.00 | 0.04 | 0.03 | 0.00 | 0.00 |
| | z | 0.00 | 0.00 | 0.00 | 0.00 | 0.00 | 0.00 |
| | | SMA | | | | | |
| | x | 1.12 | 0.00 | 0.00 | 0.00 | 0.03 | 0.03 |
| | y | 0.00 | 1.12 | 0.16 | 0.02 | 0.00 | 0.00 |
| | z | 0.00 | 0.00 | 0.00 | 0.03 | 0.00 | 0.00 |

| | | ISU | | | | | |
|---|---|---|---|---|---|---|---|
| | | x | y | z | θx | θy | θz |
| Exit Pupil Centre (mm) | x | 0.00 | 0.00 | 0.00 | 0.00 | 2.92 | 0.00 |
| | y | 0.00 | 0.06 | 0.00 | 2.98 | 0.00 | 0.00 |
| | | ISA | | | | | |
| | x | 27.85 | 0.00 | 0.00 | 0.00 | 6.63 | 0.73 |
| | y | 0.01 | 28.09 | 6.47 | 6.78 | 0.00 | 0.00 |
| | | PMA | | | | | |
| | x | 0.43 | 0.00 | 0.00 | 0.00 | 0.39 | 0.02 |
| | y | 0.00 | 0.65 | 8.64 | 0.53 | 0.00 | 0.00 |
| | | SMA | | | | | |
| | x | 28.28 | 0.00 | 0.00 | 0.00 | 0.87 | 0.80 |
| | y | 0.01 | 28.65 | 2.22 | 0.38 | 0.00 | 0.00 |

In conclusion, due to the exit pupil requirement, the sensitivity coefficients on ISU elements are very stringent. To relax MAIT tolerances, we can try to rebalance our budget, or to use a compensator. That will be implemented in the next phase if necessary.

## 5. CONCLUSION

Since Harmoni Interim Study in 2014, a big effort from CRAL has been made to pay attention onto non-compliances which were cost, delivery time and feasibility. CRAL team has managed to propose an affordable ISM opto-mechanical design. This new ISM optical design satisfies most of requirements including image quality, pupil and slit position. Nevertheless, the analysis presented here shows that the design of Harmoni ISM is compliant but without margins for MAIT Phase especially for the output pupil when the Image Slicer is combined with the Relay System Module [2]. For the Final Design Review, CRAL team will improve the margins which will ease the MAIT Phase for example in using compensators.

In conclusion, the ISM optical design still provides expected performances with small margins in the critical areas of output pupil positioning. There is then a good hope that the finally build ISM will match its ambitious scientific performances.